\documentclass{aa}
\usepackage{amsmath}
\usepackage{graphicx}
\usepackage{natbib}
\usepackage{txfonts}
\bibpunct{(}{)}{;}{a}{}{,}
\newcommand{\ha}{H$\alpha$}

\newcommand{\kpc}{{\rm kpc}}
\newcommand{\myr}{\,$M_{\sun}\,{\rm yr}^{-1}$}
\newcommand{\ro}{\,$R_{\sun}$}

\newcommand{\lo}{\,$L_{\sun}$}
\newcommand{\cmd}{\,cm$^{-2}$}
\newcommand{\cmt}{\,cm$^{-3}$}
\newcommand{\ecs}{$\rm\,erg\,cm^{-2}\,s^{-1}$}
\newcommand{\ecsa}{$\rm\,erg\,cm^{-2}\,s^{-1}\,\AA^{-1}$}

\begin{document}

%
%
%

\title{The origin of the supersoft X-ray--optical/UV flux 
                     anticorrelation\\ 
                in the symbiotic binary AG~Draconis}  

\author{A.~Skopal
        \inst{1}
        \and
        M.~Seker\'a\v{s}
        \inst{1}
        \and
        R.~Gonz\'alez-Riestra
        \inst{2}
        \and
        R. F.~Viotti
        \inst{3}
}
\institute{Astronomical Institute, Slovak Academy of Sciences,
           059\,60 Tatransk\'{a} Lomnica, Slovakia 
       \and
           XMM Science Operations Centre, ESAC, PO Box 78, 
           28691 Villanueva de la Ca\~nada, Madrid, Spain
       \and
           INAF Instituto di Astrofisica Spaziale e Fisica Cosmica 
           di Roma, via del Fosso del Cavaliere 100, 00133 Roma, Italy
}
\date{Received / Accepted}

\abstract
 {AG~Draconis produces a strong supersoft X-ray emission. 
  The X-ray and optical/UV fluxes are in a strict 
  anticorrelation throughout the active and quiescent phases. 
 }
 {To identify the source of the X-ray emission and reveal 
  the nature of the observed flux anticorrelation.
 }
 {The X-ray and UV observations with \textsl{XMM-Newton}, far-UV 
  spectroscopy from \textsl{FUSE}, low- and high-resolution 
  \textsl{IUE} spectra 
  and optical/near-IR spectroscopic and/or photometric 
  observations. Modeling the spectral energy distribution and 
  broad wings of the \ion{O}{vi}~$\lambda 1032, \lambda 1038$ 
  and \ion{He}{ii}\,$\lambda$1640 lines by 
  the electron-scattering during the maximum of the 2003 burst, 
  the following transition and quiescent phase. 
 }
 {The X-ray--near-IR energy distribution at different levels 
  of the star's brightness confirmed quantitatively the observed 
  flux anticorrelation and showed that the optical bursts are 
  associated to an increase of the nebular component of radiation. 
  The profile-fitting analysis revealed a significant increase 
  in the mean particle density around the hot star from 
  $\sim 2.6\times 10^{10}$\cmt\ during quiescent phase to 
  $\sim 1.1\times 10^{12}$\cmt\ during the burst. 
 }
 {The supersoft X-ray emission is produced by the white dwarf 
  photosphere. The X-ray and far-UV fluxes make it possible 
  to determine its temperature unambiguously. 
  The supersoft X-ray--optical/UV flux anticorrelation is caused 
  by the variable wind from the hot star. 
%
%
  The enhanced hot star wind gives rise to the optical bursts 
  by reprocessing high-energy photons from the Lyman continuum 
  to the optical/UV. 
}
\maketitle

\section{Introduction}

Symbiotic stars are long-period (orbital periods are in order 
of years) interacting binaries consisting of a cool giant and 
a compact star, which is in most cases a white dwarf (WD), 
that accretes from the giant's wind. 
%
%
This process generates a very hot ($T_{\rm h} \approx 10^5$\,K) 
and luminous ($L_{\rm h} \approx 10^2 - 10^4$\lo) source of 
radiation, that ionizes a fraction of the neutral wind from 
the giant giving rise to {\em nebular} emission. As a result 
the spectrum of symbiotic stars consists of three basic 
components of radiation -- two stellar and one nebular. 
\citep[see e.g.][]{stb,kw84,m+91,sk05}. 
This situation represents the so-called {\em quiescent phase}, 
during which the symbiotic system releases its energy approximately 
at a constant rate and at a stable spectral energy 
distribution (SED). 
%
%
Sometimes, the symbiotic system changes its radiation significantly, 
brightens up by 1--3 magnitudes in the optical, shows signatures 
of a mass-outflow and changes its ionization structure for a few 
months to years 
\citep[see e.g.][ and references therein]{viotti+84,cmm03,sk05}. 
We name this stage as the {\em active phase}, and the corresponding 
brightening in the light curve is classified as the 'Z~And-type' 
outburst. Occurrence of these outbursts is unpredictable and 
their nature is so far poorly understood 
\citep[e.g. Sect.~1 in][]{sok+06}. 
Investigation of this type of outbursts represents 
the key problem in the research of symbiotic stars. 

AG~Dra is a yellow symbiotic binary comprising a K2\,III giant 
\citep{ms99} and a WD accreting from the giant's wind on 
a 549-day orbit \citep{f+00}. The light curve of AG~Dra shows 
numerous bursts with amplitude of 1--3\,mag in $U$ (Fig.~1). 
\cite{gon+99} identified {\em cool} and {\em hot} outbursts 
differing in their Zanstra temperatures and the light curve 
profile. The former are more pronounced 
($\Delta U\sim 3$\,mag) lasting for 1--2 years 
(e.g. 1981-83, 2006-07), while the latter are weaker 
($\Delta U\sim 1-2$\,mag), single brightenings lasting for 
weeks to months (e.g. 1985, 1986, 2003, see Fig.~1). 
%
%
Modeling the UV/IR continuum, \cite{sk05} found a significant 
contribution from the nebula in the near-UV/optical that 
strengthenes during outbursts. 
%
The symbiotic nebula in AG~Dra is dense. Material supplied by 
the giant's wind into the binary environment at a rate of
$3\times 10^{-7}$\myr\ \citep[][]{sk05} corresponds to number 
densities of $\approx 1-10\times 10^{8}$\cmt\ in between the binary 
components and the neighbouring regions of similar dimensions. 
Therefore, to investigate properties of symbiotic nebula 
we can consider the Case $B$ for its radiation 
\citep[see][]{ost89,kwok00}. 
%
%

AG~Dra is a halo binary system, placed at a galactic latitude 
of +41$\degr$ with a low reddening. \cite{viotti+83} found 
$E_{\rm B-V} = 0.06\pm 0.03$ by fitting the 2\,200\,\AA\ 
depression on the \textsl{IUE} spectra. \cite{mika+95} 
re-analyzed a large number of \textsl{IUE} spectra and suggested 
$E_{\rm B-V} = 0.05\pm 0.01$, while \cite{bi+00} derived 
the extinction value of $E_{\rm B-V} = 0.08\pm 0.01$\,mag by 
fitting the UV spectra obtained by the Hopkins Ultraviolet 
Telescope. These quantities are comparable to the total 
reddening through the Galaxy towards AG~Dra 
\citep[0.07--0.08,][]{b+h82}, which justifies that the measured 
extinction to AG~Dra is purely of the interstellar nature. 
From the Ly$\alpha$ width on high-resolution \textsl{IUE} spectra, 
\cite{viotti+83} estimated the \ion{H}{i} column density of 
$\log N_{\rm H} = 20.2$. Using the Einstein X-ray observations, 
\cite{anderson+81} found $N_{\rm H} = 3\times 10^{20}$\cmd. 
According to the relationship between $N_{\rm H}$ and 
$E_{\rm B-V}$ \citep[$N_{\rm H}/E_{\rm B-V}\dot = 4.93\times 
10^{21}{\rm cm^{-2}mag^{-1}}$,][]{d+s94}, both these parameters 
are consistent. 
%
%
\begin{table*}[p!t]
\caption[]{Log of the used spectroscopic observations}
\begin{center}
\begin{tabular}{cccccc}
\hline
\hline
    Date   & Julian date  & Stage$^{\star}$ & Region  & Observatory 
           & $T_{\rm exp}$\\
YYYY/MM/DD & JD~2\,4...   &                 &  [nm]   &            
           & [s] \\
\hline
2003/10/06 & 52919.5 & HB & 330 - 1020 &\textsl{Loiano}     &~~660 \\
2003/10/10 & 52923.7 & HB & 212        &\textsl{XMM-OM}     &~1200  \\
2003/10/10 & 52923.7 & HB & 1.93 - 6.92&\textsl{XMM-Newton} &17121$^\dagger$\\
1985/03/13 & 46137.8 & HB & 115 - 335  &\textsl{IUE}        &~~600 \\
1995/07/28 & 49927.5 & HB & \ion{He}{ii}\,164.0&\textsl{IUE}&~5400 \\
2003/11/14 & 52958.5 & T  & 100 - 108.2&\textsl{FUSE}       &~4863 \\
2003/11/19 & 52963.7 & T  & 1.93 - 6.92&\textsl{XMM-Newton} &~9217 \\
2003/11/19 & 52963.7 & T  & 212 - 291  &\textsl{XMM-OM}     &~4400 \\
2004/06/15 & 53172.0 & Q  & 3.12 - 6.92&\textsl{XMM-Newton} &12354 \\
2004/06/15 & 53172.0 & Q  & 212 - 291  &\textsl{XMM-OM}     &~4400 \\
2004/06/24 & 53181.0 & Q  & 100 - 108.2&\textsl{FUSE}       &10832 \\
1980/05/23 & 44383.1 & Q  & 115 - 335  &\textsl{IUE}        &~2520 \\
\hline
\end{tabular}
\end{center}
$^{\star}$~hot burst (HB), transition to quiescence (T), 
           quiescent phase (Q), ~
           $^{\dagger}$\,no detection of supersoft X-rays
\end{table*}
%
%

The orbital inclination of the binary is rather low. There are 
no signs of eclipses either in the optical, far-UV or X-ray 
regions \citep[e.g.][]{gon+08}. 
Considering geometry of the main sources of radiation, 
\cite{mika+95} estimated a system inclination 
$i \approx 30^{\circ}-45^{\circ}$ and \cite{ss97}, based 
on spectropolarimetric observations, suggested 
$i=60\pm 8^{\circ}.2$. 
As a result we see the hot star rather from its pole than 
the orbital plane, where a disk-like structured material 
can be expected \citep[e.g.][]{mm98}. 
Both the low interstellar absorption/extinction and the low 
orbital inclination suggest a high transmission of the 
interstellar medium (ISM) to soft X-rays, which is consistent 
with the fact that AG~Dra is the brightest system at these 
energies among other symbiotics. Therefore, AG~Dra has been 
frequently observed by X-ray satellites. 

Using the ROSAT observations \cite{greiner+97} first noted 
a remarkable decrease of the X-ray flux during the optical 
(1994 and 1995) maxima, while in the UV and the optical they 
indicated a large increase of the emission line and continuum 
fluxes. They ascribed this X-ray/UV flux anticorrelation 
to a temperature decrease of the hot component. 
The X-ray history of AG~Dra was recently reviewed and discussed 
by \cite{gon+08}. They found that the anticorrelation between 
the X-ray and optical/UV emission appears to be a general feature 
of AG~Dra radiation and is independent of the type and strength 
of the outburst. They suggested that during outbursts the WD 
radiation increases, but is strongly absorbed by the circumstellar 
ionized gas. 
%

As the effect of the flux anticorrelation is related to 
transitions between active and quiescent phases, its 
understanding thus can aid us in revealing the nature 
of the Z~And-type outbursts. 
%
Accordingly, in this paper we investigate the origin of 
the observed anticorrelation between the supersoft X-ray and 
the optical/UV fluxes for the case of AG~Dra. For this 
purpose we model its composite continuum within the 
X-rays -- near-IR domain at different levels of the activity. 
In Sect.~2 we summarize and describe nearly-simultaneous 
observations we used to model the continuum. 
In Sect.~3 we describe our analysis and present the results. 
Their discussion and summary with conclusions are found in 
Sects.~4 and 5, respectively. 

%
%
%
%
%
\begin{figure*}
\centering
\begin{center}
\resizebox{\hsize}{!}{\includegraphics[angle=-90]{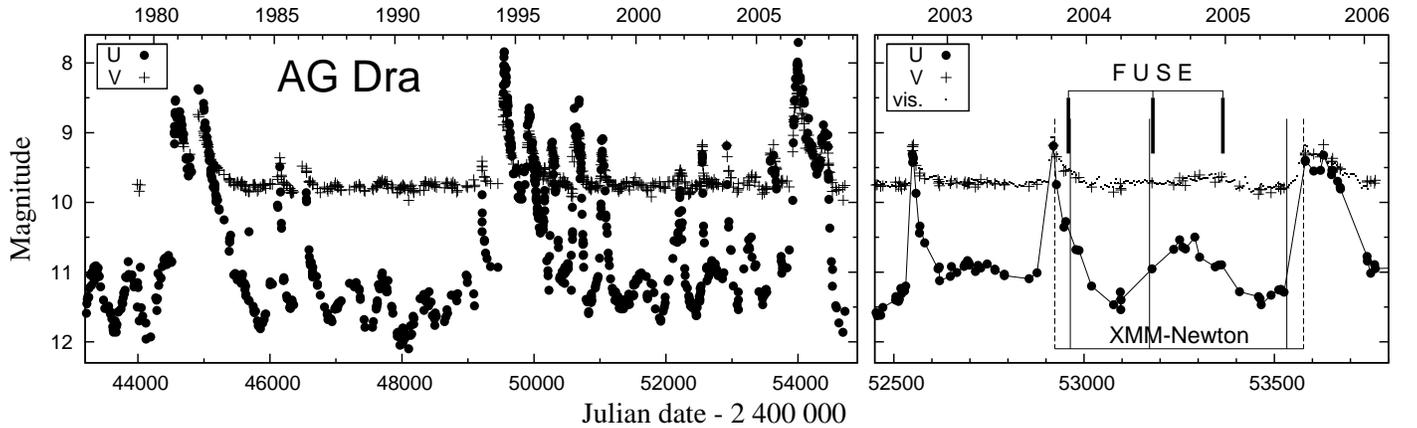}}
%
\caption[]{
The $U$ and $V$ light curves of AG~Dra from 1977. They 
are characterized with a series of outbursts with multiple 
maxima. 
The data were summarized by \cite{sk+07}. 
The right panel shows a detail around the 2003 hot burst 
with timing of the \textsl{XMM-Newton} (long thin lines) and 
the \textsl{FUSE} 
(short thick bars) observations. During the optical maxima, 
the soft X-ray emission was not detectable (long dashed lines). 
          }
\end{center}
\end{figure*} 

\section{Observations}

\subsection{Sources of the data and their timing}

For the aim of this paper we selected observations from 
the supersoft X-ray to the near-IR, taken at different 
levels of the AG~Dra activity. 
%
We analyze the archival supersoft X-ray data made by 
the X-ray Multi-Mirror Telescope (\textsl{XMM-Newton}), 
including its Optical Monitor (\textsl{XMM-OM}) for the near-UV 
fluxes, the far-UV spectra (988 -- 1082\,\AA) made with the 
LiF1A channel of the Far Ultraviolet Spectroscopic Explorer 
(\textsl{FUSE}), the ultraviolet low- and high-resolution spectra 
taken by the International Ultraviolet Explorer (\textsl{IUE}), 
the optical low-resolution spectrum from the Loiano observatory 
and the flux-points determined by the broad band optical 
$UBVR_CI_C$ and near-IR $JHKLM$ photometry. 
The \textsl{XMM-Newton} and Loiano observations were described in 
detail by \cite{gon+08}. The \textsl{FUSE} spectra were processed 
according to \cite{sk+06}. 

During quiescent phase we composed the observed SED by 
the \textsl{XMM-Newton} and \textsl{FUSE} observations, 
complemented with the \textsl{IUE} spectrum (SWP9084/LWR07831) 
taken at a similar orbital position. 
During transition phase we selected the near-simultaneous 
\textsl{XMM-Newton} and \textsl{FUSE} observations and the 
photometric $UBV$ flux-points from the descending branch of 
the 2003 burst (Fig.~1). 
To model the SED at the maximum of hot bursts we used the 
ultraviolet \textsl{XMM-OM} fluxes and the optical spectrum, 
both from the maximum of the 2003 burst. However, to estimate 
the relevant far-UV fluxes we had to use a non-simultaneous 
observation. We used well-exposed \textsl{IUE} spectra 
(SWP25443/LWP05513) taken around the maximum of the 1985 hot 
burst. To match the \textsl{XMM-OM} fluxes we scaled 
the \textsl{IUE} spectrum by a factor of 1.3. 
This was possible, because hot bursts are similar in their 
profiles and colours \citep[see Fig.~1 here and Fig.~6 of][ for 
UV colours]{gon+99}. Their SED shows a dominant contribution from 
the nebula to the near-UV/U region, while the hot stellar source 
dominates the far-UV region \citep[see Fig.~13 of][]{sk05}. 

For the profile-fitting analysis we used the 
\ion{O}{vi}~$\lambda 1032, \lambda 1038$ doublet on the \textsl{FUSE} 
spectra, available from the quiescent and transition phases. 
For the optical maximum we analyzed the 
\ion{He}{ii}\,$\lambda 1640$ line exposed on the high-resolution 
\textsl{IUE} spectra (SWP55372 and SWP55373) at the maximum of 
the 1995 hot burst. The spectra were calibrated with the aid 
of their low-resolution counterparts. 

Relevant observations were dereddened with $E_{\rm B-V}$ = 0.08 
and resulting parameters were scaled to a distance of 1.1\,kpc 
\citep[][]{bi+00, sk05}. Their log and plots are given in 
Table~1 and Fig.~2, respectively. 

\subsection{Derivation of the X-ray fluxes}

In this paper we analyze observations of AG~Dra made by 
\textsl{XMM-Newton} \citep[][]{jansen}, performed during 
the 2003 hot burst (at its optical maximum and transition 
to quiescence) and during quiescent phase (Fig.~1). 
Details of the observations and the data reduction were 
already described by \cite{gon+08}. 
To estimate the X-ray fluxes we fitted the spectra from 
the EPIC-pn instrument with the XSPEC software package. 
According to a very high temperature of the hot stellar source in 
AG~Dra \citep[$> 10^{5}$\,K,][]{kw84,m+91,greiner+97,sk05} we can 
assume that the supersoft X-ray fluxes are emitted by the WD 
photosphere. Therefore, we considered a simple temperature 
blackbody model in fitting our 0.18--0.4\,keV data. 
The resulting fit to the spectrum from quiescence corresponded 
to the blackbody radiation absorbed with the hydrogen column 
density $N_{\rm H} \sim 3\times 10^{20}$\cmd and temperature 
of $\sim$14.5\,eV as in the case of the ROSAT PSPC observations 
from quiescence \citep[see][]{greiner+97,gon+08}. 
During the transition, the shape of the spectrum was similar, 
but the lower signal to noise ratio prevented a meaningful fit. 
In this case we assumed the same parameters as for the 
quiescence spectrum, and thus fitted only the normalization 
factor.
%
Fluxes were derived from these 'unfolded' models and 
the corresponding data/model ratio. This approach implies 
that the fluxes obtained by this way are, to a certain extent, 
model dependent. However, we are confident that the spectrum 
can be well represented by a simple absorbed blackbody model, 
which is also supported by the data-minus-model residuals 
that are small, flat and within the errors in the range of 
40--70\,\AA. Here we refer the reader to the paper of 
\cite{nowak+02}, who discuss other approaches to obtaining 
flux points from X-ray data and the effects connected with. 
To assess influence of the hydrogen column density and the 
temperature in the derived fluxes we fitted the data with 
a few hundred of models covering a wide range of these parameters. 
Taking the 30 best fit cases (in terms of $\chi^2$), we found 
that the fluxes derived from these models differ from 
the resulting ones by less than 15\% in the range of 30-65\,\AA. 
Uncertainties of the X-ray fluxes derived in this way satisfy 
the purposes of this work. 
Further, we complemented the X-ray fluxes with those from 
the \textsl{FUSE} spectra to determine final parameters of 
the radiation produced by the WD photosphere 
(see below, Sect.~3.1.1.). 

\section{Analysis and results}

\subsection{Modeling the composite spectrum}

The continuum spectrum of symbiotic stars is composed 
of three basic components of radiation -- two stellar, 
$F_{\rm h}(\lambda)$ and $F_{\rm g}(\lambda)$, from the hot 
star and the cool giant, respectively, and one nebular, 
$F_{\rm n}(\lambda)$, from the ionized gas in the system 
(Sect.~1). 
Their superposition then yields the observed flux as 
%
%
\begin{equation}
  F(\lambda) =  F_{\rm g}(\lambda) + F_{\rm h}(\lambda) +
                F_{\rm n}(\lambda).
\end{equation}
%
To achieve our aim we need to determine physical parameters 
of the hot stellar and the nebular component of radiation during 
different stages of the activity. 
We will use the method of disentangling the composite spectrum 
of S-type symbiotic stars as proposed by \cite{sk05}. 
Observations from the soft X-rays to the near-IR cover well 
all the energy domains, within which the individual components 
of radiation have a dominant contribution. This allows us 
to model them separately. 
In the following three sections we describe our approximations 
of these components of radiation. 
%

\subsubsection{The hot star continuum}

According to \cite{greiner+97} and \cite{viotti+05} the soft 
X-ray radiation in AG~Dra could be associated with the hot 
star photosphere. 
However, the soft X-rays from cosmic sources are significantly 
attenuated by absorptions in the ISM, which requires relevant 
correction before their interpretation 
\citep[e.g.][]{crudd+74,wilms+00}. Assuming that no emission 
occurs on the path between the X-ray source and the observer, 
we can use the simplest solution of the radiative transfer 
equation to correct the observed X-ray flux, 
$F_{\rm x}^{\rm obs}(\lambda)$, for absorptions as 
%
%
\begin{equation}
 F_{\rm x}^{\rm obs}(\lambda) = 
           F_{\rm x}(\lambda)\,e^{-\tau_{\lambda}},
\end{equation}
where $F_{\rm x}(\lambda)$ is the X-ray spectrum as emitted 
by the source and $\tau_{\lambda}$ is the optical depth 
along the line of sight. It is given by the absorption 
cross-section per atom, $\sigma_i(\lambda)$, of the element, 
$i$, and the total number of atoms on the line 
of sight, $\sum_i n_i$. Thus, 
$\tau_{\lambda} = \sum_i n_i \sigma_i(\lambda) = 
\sum_i a_i\sigma_i(\lambda) N_{\rm H} = 
\sigma_{\rm X}(\lambda) N_{\rm H}$, where $a_i$ is the 
relative abundance, $N_{\rm H}$ is the total hydrogen column 
density [\cmd] and $\sigma_{\rm X}(\lambda)$ [cm$^{2}$] is 
the total cross-section of the material on the line of sight 
per hydrogen atom in the X-ray domain \citep[e.g.][]{crudd+74}. 

In addition to the X-ray fluxes from \textsl{XMM-Newton}, 
we use the far-UV continuum fluxes between $\lambda$1188 
and $\lambda$1000\,\AA, made nearly-simultaneously with 
the \textsl{FUSE} satellite (see Fig.~1). 
As we analyze both the X-ray and the far-UV data, we consider 
attenuation of the light by bound-free absorptions in 
the X-ray domain and by the interstellar extinction in 
the far-UV spectrum. 
According to previous models (see Sect.~2.2.), we approximate 
the hot stellar continuum by a blackbody radiation at 
a temperature $T_{\rm h}$. 
As a result, and in the sense of Eq.~(2), we model the hot 
star continuum by fitting the observed X-ray/far-UV fluxes, 
$F_{\rm h}^{\rm obs}(\lambda)$, with a function 
%
%
\begin{equation}
        F_{\rm h}^{\rm obs}(\lambda) =
 \begin{cases}
       \theta_{\rm h}^2 \pi B_{\lambda}(T_{\rm h})\,
       e^{-\sigma_{\rm X}(\lambda)\,N_{\rm H}}
     & \text{for } \lambda < 912\,\AA \\[2mm]
       \theta_{\rm h}^2 \pi B_{\lambda}(T_{\rm h})\,
       10^{-0.4\,R\,k_{\lambda}\,E_{\rm B-V}} 
     & \text{for } \lambda > 912\,\AA
 \end{cases}
\end{equation}
%
%
where the scaling factor $\theta_{\rm h}=R_{\rm h}^{\rm eff}/d$ 
is given by its effective radius $R_{\rm h}^{\rm eff}$ and 
the distance $d$, and thus represents the angular radius of 
the hot stellar source. 
The observed far-UV fluxes were corrected using the extinction
curve $k_{\lambda}$ of \cite{c+89} and the ratio of total to
selective extinction $R = 3.1$ \citep[e.g.][]{wegner}.
In addition to the attenuation effects included in Eq.~(3), 
the neutral hydrogen on the line of sight causes a depression 
of the continuum around hydrogen lines of the Lyman series due 
to the Rayleigh scattering \citep[][]{inv89}. 
By analogy to the first term of Eq.~(3) we can express 
the Rayleigh attenuated continuum as 
%
%
\begin{equation}
 F_{\rm h}^{\rm obs}(\lambda) = 
       \theta_{\rm h}^2 \pi B_{\lambda}(T_{\rm h})\,
       e^{-\sigma_{\rm R}(\lambda)\,N_{\rm H}}, 
\end{equation}
where $\sigma_{\rm R}(\lambda)$ is the Rayleigh scattering 
cross-section for atomic hydrogen. On our \textsl{FUSE} 
spectra the effect is seen around the Ly-$\beta$ line 
(Fig.~2), and can be used to test quantity of $N_{\rm H}$ 
obtained from the X-ray domain. 
%
%
%
\begin{figure*}
\centering
\begin{center}
\resizebox{\hsize}{!}{\includegraphics[angle=-90]{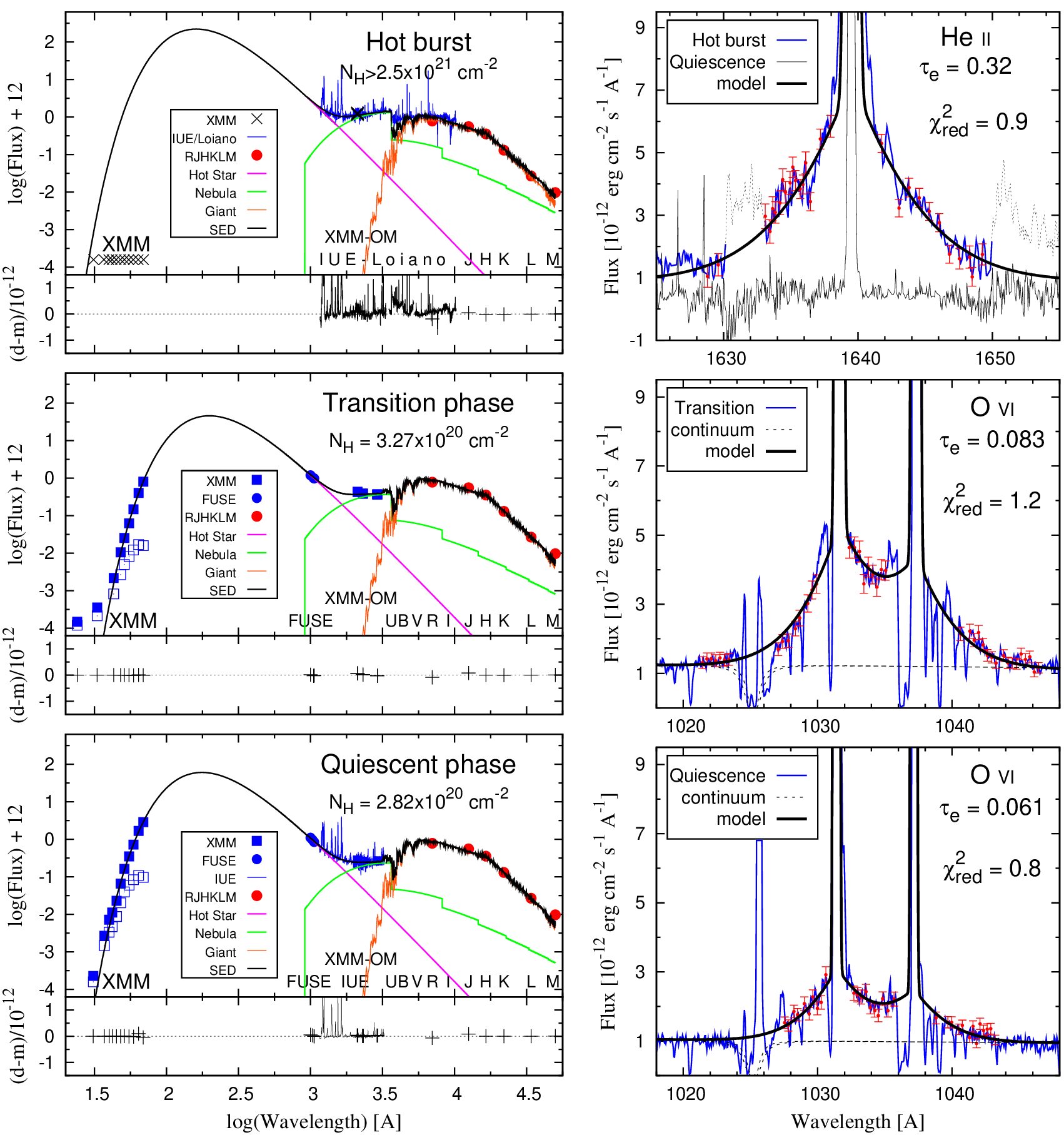}}
%
\caption[]{
Left panels show a comparison of the observed and modeled SEDs 
of AG~Dra during the hot burst (top), transition phase (middle) 
and quiescent phase (bottom) with corresponding residuals. 
Open/filled squares are the observed/corrected X-ray fluxes. 
They are in units of $10^{-12}$\ecsa. Typical uncertainties 
of the flux-points are around of 10\% (Sect.~3.3). 
Right panels compare the observed and modeled broad wings of 
the \ion{O}{vi}~$\lambda 1032, \lambda 1038$ doublet and the 
\ion{He}{ii}~$\lambda 1640$ line at these stages of activity. 
Sample points with uncertainties are plotted in red. Timing 
of individual observations are given in Table~1. Models are 
described in Sect.~3 and their parameters are given 
in Tables~2 and 3. 
          }
\end{center}
\end{figure*}

\subsubsection{The nebular continuum}

We approximate the nebular component of radiation in the 
UV/optical continuum by processes of recombination and thermal
bremsstrahlung in the hydrogen plasma, radiating under 
conditions of the Case $B$. Validity of this simplification 
is supported by the following arguments: 
  (i) A strong nebular continuum is characterized with 
a simple and low electron temperature of 
$T_{\rm e} \sim 2\times 10^{4}$\,K (Table~3), which suggests 
that the nebula is powered mainly by photoionization, 
i.e. ionizations by collisions are not important. 
  (ii) There are no recognizable signatures of the \ion{He}{ii} 
continuum in the \textsl{IUE} spectra (e.g. a jump in emission 
at $\sim$2050\,\AA\ and/or a pronounced Paschen series of the 
\ion{He}{ii} recombination lines). Also it is not possible 
to separate contributions from the \ion{He}{i} continuum, 
because of its very similar recombination coefficients to 
that of \ion{H}{i} \citep[e.g.][]{b+m70} and a small abundance. 
That is why we consider nebular emission from hydrogen only. 
  (iii) Symbiotic nebulae are relatively very dense (Sect.~1). 
This makes the mean-free path of any diffuse Lyman continuum 
photons to be too short to escape the nebula, i.e. the nebula 
is optically thick in the Lyman continuum. This means that 
the ionizations caused by stellar radiation-field photons 
are balanced by recombinations to excited levels of \ion{H}{i}, 
while the ground state in the recombination process can be 
ignored. Therefore we consider the Case $B$ for the nebular 
radiation. 
  (iv) Because of the high density, we can neglect contributions 
due to the two-photon emission. 

According to these simplifications and with the aid of Eq.~(11) 
in \cite{sk05}, the $F_{\rm n}(\lambda)$ term in Eq.~(1) here 
can be expressed as 
%
%
\begin{equation}
F_{\rm n}(\lambda) = \frac{EM}{4\pi d^2}
      \varepsilon_{\lambda}({\rm H},T_{\rm e}),
\end{equation}
%
where $\varepsilon_{\lambda}({\rm H},T_{\rm e})$ 
[$\rm\,erg\,cm^{3}\,s^{-1}\,\AA^{-1}$] is the volume 
emission coefficient for hydrogen, which depends on 
the electron temperature $T_{\rm e}$ and is a function 
of wavelength \citep[e.g.][]{b+m70}. We note that the observed 
nebular emission represents only the optically thin part of 
the symbiotic nebula \citep[][]{sk01}. This implies 
that the measured $EM$ represents a lower limit of that 
originally created by ionizations, which puts a lower 
limit to the required flux of ionizing photons 
(see Eq.~(6) below). 

\subsubsection{Radiation from the giant}

For the stellar radiation from the giant in AG~Dra we adopt 
the model SED according to \cite{sk05}. The model is based 
on the optical $VRI$ and the near-IR $JHKLM$ flux points 
matched by a synthetic spectrum calculated for the effective 
temperature of 4300\,K. This spectrum then defines 
the first term in Eq.~(1), $F_{\rm g}(\lambda)$. 
Its bolometric flux, 
$\theta_{\rm g}^2 \sigma T_{\rm eff}^4 = 
9.51\times 10^{-9}$\,\ecs, corresponds to the giant's 
radius 
  $R_{\rm g} = 33\,(d/1.1\,\kpc)\,R_{\sun}$ 
and the luminosity 
  $L_{\rm g} = 360\,(d/1.1\,\kpc)^2\,L_{\sun}$. 

\subsection{Thomson-scattering wings of 
            \ion{O}{vi}\,$\lambda 1032,\lambda 1038$ 
            and \ion{He}{ii}\,$\lambda 1640$ lines}

The aim of this section is to model the extended wings of 
the \ion{O}{vi} and \ion{He}{ii} lines measured at similar 
levels of the optical brightness, at which the X-ray 
observations were carried out. By this way we support the 
results obtained by model SEDs. 

\cite{schmid+99} suggested that the broad wings of the 
\ion{O}{vi}~$\lambda$1032 and $\lambda$1038 resonance lines 
could be explained by scattering of the \ion{O}{vi} photons 
by free electrons. The effect of this process is weak and 
wavelength independent, because of a very small and 
constant value of the Thomson cross-section, 
$\sigma_{\rm T} = 6.652\times 10^{-25}$\,cm$^{2}$. 
However, the densest portions of the symbiotic nebula 
($\log(n_{\rm e}) \sim 8 - 12$\,\cmt) extending to a few AU 
could be sufficiently optically thick for 
the electron-scattering process. 
%
%
Qualitatively, the effect should be more significant during 
active phases, because of a surplus of electrons from the 
increased hot star wind \citep[e.g.][]{sk06}. Especially, 
the strong emission lines of highly ionized elements that 
are formed in the densest part of the hot stellar wind 
represent the best candidates for a well measurable effect 
of the Thomson scattering. 

To calculate the electron-scattering wing profiles we 
adopted the procedure suggested by \cite{munch50} that 
assumes the electron scattering happens in the layer 
outside the line formation region, and that the electrons 
are segregated from the other opacity sources, which implies 
no change in the equivalent width of the line. Further we 
used the expression of \cite{castor+70} for the resulting 
scattered line profile and the electron-scattering 
redistribution function derived by \cite{h+m67}. 
This simplified approach was used by many authors, recently 
by \cite{young+05}. 

In our profile-fitting analysis, we approximated the observed 
P~Cygni-type of the \ion{O}{vi} line profiles by two Gaussians 
and fitted their broad wings with 2 free 
parameters -- the electron temperature, $T_{\rm e}$, and 
the electron-scattering optical depth, $\tau_{\rm e}$. 
Removing emission/absorption features and bumps in the profiles 
we fitted the broad wings in the range of about $\pm 10$\,\AA\
around the line cores using 644, 1004 and 370 flux-points 
in the wing profile from quiescence, transition and burst 
stage, respectively. 
The resulting fits for all three cases we analyzed have 
small reduced $\chi^2_{\rm red} \la 1$ (Table~2, Fig.~2), 
which implies that mean residuals are comparable to the errors 
of the modeled flux-points. Thus the model fits well the broad 
wings of the profile, which confirms their origin as due to 
the electron-scattering. 
A detailed description of our approach will be presented 
elsewhere \citep[][]{se+sk}. 

A comparison of our model SEDs and profiles with observations 
is plotted in Fig.~2 and corresponding parameters are given 
in Tables~2 and 3. 
%
%
\begin{table}
\caption[]{Parameters of the profile-fitting analysis (Sect.~3.2, Fig.~2)}
\begin{center}
\begin{tabular}{ccccccc}
\hline
\hline
   Date   & Stage& Line &$\tau_{\rm e}$ &$T_{\rm e}$&$\chi^2_{\rm red}$&
$\bar{n}_{\rm e}$\\
           &      &      &               &    [K]   &                  & 
[cm$^{-3}]$      \\
\hline
1995/07/28 & HB& \ion{He}{ii} & 0.32~ & 28\,800&0.9&$1.1\times 10^{12}$\\
2003/11/14 & T & \ion{O}{vi}  & 0.083 & 28\,500&1.2&$1.7\times 10^{11}$\\
2004/06/24 & Q & \ion{O}{vi}  & 0.061 & 20\,300&0.8&$2.6\times 10^{10}$\\
\hline
\end{tabular}
\end{center}
\end{table}
\subsection{The SED-fitting analysis}

\subsubsection{Parameters of the hot stellar source}

To derive parameters of the hot stellar source in AG~Dra, 
the following points are relevant. 

(i) We dereddened the \textsl{FUSE} fluxes and calculated
the Rayleigh scattering effect around Ly-$\beta$ (Eq.~(4))
to estimate $N_{\rm H}$. The scattering cross-section,
$\sigma_{\rm R}(\lambda)$, was calculated according to
\cite{nsv89}. 
However, numerous and strong absorptions in the \textsl{FUSE} 
spectrum did not allow us to determine unambiguously 
the $N_{\rm H}$ parameter. Therefore, we could only compare 
the Rayleigh attenuated blackbody radiation to the observed 
continuum for a reasonable range of $N_{\rm H}$ quantities.
We found that the observed depression in the continuum
around Ly-$\beta$ constrains $N_{\rm H}$ to
$\sim 2-4\times 10^{20}$\,\cmd\ that is consistent with
the values suggested by different methods (Sect.~1). 

(ii) We calculated a grid of synthetic models of the 
function (3) for reasonable ranges of the fitting parameters, 
$\theta_{\rm h}$, $T_{\rm h}$ and $N_{\rm H}$. We compared 
models to both the observed X-ray and dereddened UV fluxes, 
and used the $\chi^{2}$ statistics to evaluate the fit. 
To correct the observed X-ray fluxes for absorptions 
we used the {\em tbabs} absorption model 
\citep[T\"ubingen-Boulder absorption ISM model,][]{wilms+00}. 
%
%
During the quiescent phase we fitted 11 X-ray fluxes from 31 
to 69\,\AA\ and three far-UV flux-points around 1000\,\AA\ 
(bottom left panel of Fig.~2). 
We estimated errors in the X-ray fluxes to be of 2---17\%
of their mid values, while for the \textsl{FUSE} spectra 
we adopted errors as large as 5---10\% of the measured 
continuum. 
The best-fit-model and the flux-point errors yielded 
$\chi^2$ = 34.6 and $\chi^2_{\rm red}$ = 3.1 
for 11 degrees of freedom (dof). As the model fits well the 
data, the somewhat larger $\chi^2_{\rm red}$ reflects rather 
small values of our error estimates. 
During the transition stage we fitted 7 X-ray fluxes between 
43 and 69\,\AA\ and three far-UV flux-points. We omitted 
fluxes around 20--30\,\AA, because of their different nature. 
They could be originated by shocks in the nebula. 
Flux uncertainties we used, were estimated to be in the range 
of 7---22\%. The resulting fit has a small 
$\chi^2_{\rm red}$ = 0.6 for 7 dof, which suggests that the model 
fits the data very closely, well within their uncertainties. 
Corresponding best-fit-model parameters and their derivatives 
(the effective radius and the luminosity of the hot stellar 
source) are introduced in Table~3. 

(iii)
During the bursts, when the supersoft X-ray emission is absorbed 
entirely, it is not possible to determine unambiguously 
the hot star temperature for $T_{\rm h} \ga 10^{5}$\,K, 
because of too small range of the far-UV fluxes available. 
Nevertheless, we can estimate a lower limit of the temperature 
of the ionizing source, $T_{\rm h}^{\rm min}$, at which the hot 
star radiation, scaled to the far-UV fluxes, is {\em just} 
capable of producing the observed $EM$, i.e. the total number 
of ionizing photons just balances the total number 
of recombinations. 
According to simplifications for the plasma radiation as 
introduced in Sect.~3.1.2., we solve equilibrium equation 
(Eq.~6 below) for the temperature $T_{\rm h}$ and the scaling 
factor $\theta_{\rm h}^2$, which determine the flux of ionizing 
photons, with the input parameters of the measured nebula, 
$EM$ and $T_{\rm e}$. The resulting temperature then 
corresponds to $T_{\rm h}^{\rm min}$, at which the hot star 
radiation gives rise to the observed nebular emission 
\citep[see][ in detail]{sk05}. 
%

\subsubsection{Parameters of the nebula}

According to Eq.~(1), the nebular component of radiation
can be obtained by subtracting the stellar contributions 
from the observed spectrum, i.e. $F_{\rm n}(\lambda) = 
F(\lambda)-F_{\rm h}(\lambda)-F_{\rm g}(\lambda)$. 
First, we estimated representative continuum fluxes at about 
20 wavelengths between 1250 and 3300\,\AA\ and complemented 
them with photometric $U$ and \textsl{XMM-OM} fluxes, if 
available. Uncertainties of the well exposed spectra from 
the \textsl{IUE} archive are between 5 and 10\% of the measured 
values. Errors in the \textsl{XMM-OM} fluxes were estimated 
to only a few percents \citep[see Table~2 in][]{gon+08}. 
%
Uncertainties of classical photometric measurements can                
be assumed to be less than 10\%. We corrected corresponding 
fluxes for the influence of emission lines using our Loiano 
spectra \cite[see][ in detail]{sk07}. 
Second, we subtracted the contribution of the WD and the 
giant and fitted the function $F_{\rm n}(\lambda)$ given 
by Eq.~(5) to the corrected flux-points to determine the 
$EM$ and $T_{\rm e}$ of the symbiotic nebula. Resulting 
fits have $\chi^2_{\rm red} \la 1$ that proves assumptions 
of our simplified model of the nebula (Sect~3.1.2.). 
Corresponding plots and parameters are in Fig.~2 and 
Table~3, respectively. 
%
%

\section{Discussion}

\subsection{Model SEDs and the flux anticorrelation}

Model SEDs show that the light variations in 
the optica/near-UV, as measured at different levels of 
activity, are caused exclusively by the variable nebular 
continuum (Fig.~2, Table~3). 
Contribution from the hot stellar object can be neglected 
within this domain, and that from the giant can be assumed 
to be constant. 
%

The nebular component of radiation represents a fraction of 
the hot stellar radiation transformed by the 
ionization/recombination events throughout the symbiotic nebula. 
The result of this process depends on the number of ionizing 
photons ($L_{\rm ph}$ [s$^{-1}$]) produced by the hot star and 
the number of particles on their path that are subject to 
ionization. 
In the case of the hydrogen plasma (see Sect.~3.1.2.) 
characterized with one $T_{\rm e}$ and a mean particle 
concentration, $\bar{n}$, the equilibrium condition 
between $L_{\rm ph}$ photons and the number of 
recombinations can be expressed as 
%
%
%
\begin{table*}[p!t]
\begin{center}
\caption{Parameters of the SED-fitting analysis 
         (see Sect.~3.3, Fig.~2). 
        }
\begin{tabular}{ccccccccccc}
\hline
\hline
Stage                &
$N_{\rm H}$          &
$T_{\rm h}$          &
$\theta_{\rm h}$     &
$R_{\rm h}^{\rm eff}$&
$L_{\rm h}$          &
$L_{\rm ph}({\rm H})$&
$\chi^2_{\rm red}$   &
$T_{\rm e}$          &
$EM$                 &
$\chi^2_{\rm red}$   \\
                     &
[cm$^{-2}$]          &
[$K$]                &
                     &
[$R_{\sun}$]         &
[$L_{\sun}$]         &
[s$^{-1}$]           &
                     &
[$K$]                &
[cm$^{-3}$]          &
                     \\
%
%
\hline
HB &$>$2.5$\times 10^{21}$ &$>$180\,000 &$<$8.9$\times10^{-13}$&$<$0.043
   &$>$1\,760&$>8.9\times 10^{46}$&--&18\,600&6.2$\times 10^{59}$& 1.3 \\
 T &3.27$\times 10^{20}$   & 146\,200   &7.3$\times 10^{-13}$  &0.036
   & 530&3.1$\times 10^{46}$&0.6& 21\,500 & 2.2$\times 10^{59}$& 0.8   \\
 Q &2.82$\times 10^{20}$   & 164\,400   &6.3$\times 10^{-13}$  &0.031
   & 630&3.4$\times 10^{46}$&3.1& 21\,000 & 1.3$\times 10^{59}$& 0.4   \\
%
%
\hline
\end{tabular}
\end{center}
\end{table*}
%

\begin{equation}
\begin{split}
  L_{\rm ph}({\rm H})\, = \!\! \int_V \! n_{\rm e}n^{+}
            \alpha_{\rm B}({\rm H},T_{\rm e})\,{\rm d}V 
   = \alpha_{\rm B}({\rm H},T_{\rm e})\,\bar{n}^2\,V    \\
   = \alpha_{\rm B}({\rm H},T_{\rm e})\,EM,
\end{split}
\end{equation}
where $\alpha_{\rm B}({\rm H},T_{\rm e})$ [cm$^{+3}$\,s$^{-1}$] 
is the recombination coefficient to all but the ground state 
of hydrogen (i.e. the Case $B$). 
%
For parameters of AG~Dra during quiescence, the equilibrium 
condition is satisfied at the locus of points in directions 
from the hot to/around the cool star, where the flux of 
$L_{\rm ph}$ photons are balanced by the flux of the neutral 
atoms of hydrogen in the wind from the giant. 
Throughout the remainder part of the nebula 
$L_{\rm ph} > \alpha_{\rm B} EM$, which means that 
a fraction of $L_{\rm ph}$ photons escapes the nebula 
without being transformed to the nebular radiation 
\citep[see Appendix B and Fig.~3 in][]{sk01}. 
Under these conditions, new particles injected into such 
the nebula will consume the surplus of ionizing photons, 
what consequently will increase the nebular emission 
observed during the bursts. This situation can naturally 
be explained by a strengthened wind from the WD, that also 
induces an increase of the optical depth in the continuum 
in the direction of the WD, what we indicate by its larger 
effective radius in comparison with values from quiescence 
(Table~3). 
%
%

According to Eq.~(2) and the relatively large values of 
$\sigma_{\rm X}$ in the supersoft X-ray domain, 
a small increase of $N_{\rm H}$ in between the source 
and the observer produces a significant attenuation 
of the $F_{\rm x}(\lambda)$ fluxes. For example, comparison 
of our \textsl{XMM-Newton} observations from the 
transition and quiescent phases shows that the increase 
of the $N_{\rm H}$ value by a factor of only 1.16 (Table~3) 
produces attenuation of the observed fluxes at 
$\sim 0.18$\,keV by a factor of $\sim 5$ (see Fig.~2). 
At the maximum of the 2003 burst, the 0.19 -- 0.4\,keV 
emission was not detectable, in spite of the increase of 
both the luminosity and the temperature of the hot source 
(Table~3, Fig.~2). 
The high limiting quantities of $L_{\rm h}$ and $T_{\rm h}$ 
are constrained by the minimum flux of photons capable 
of ionizing hydrogen, 
  $L_{\rm ph}({\rm H}) = 8.9\times 10^{46}$\,s$^{-1}$, 
that is required to give rise the observed high amount of 
the emission measure, 
  $EM = 6.2\times 10^{59}(d/1.1{\rm kpc})^2$\cmt\ (see Eq.~(6)). 
%
%
We found that the value of 
$N_{\rm H} \ga 2.5\times 10^{21}$\cmd\ is sufficient to damp 
down the model fluxes ($\approx 5.5\times 10^{-10}$\ecs\ 
within the 0.19--0.4\,keV range) below a detection level 
of around $10^{-16}$\ecsa\ for the exposure time used 
by \textsl{XMM-Newton} on 2003/10/10. 

According to the model SEDs we ascribe the observed flux 
anticorrelation to variable wind from the hot star. 
The wind particles enrich the plasma surrounding the WD's 
photosphere, which increases the number of both the bound-free 
absorptions (parametrized by $N_{\rm H}$ in the model) 
and the free-bound emissions. 
The former {\em attenuates} the supersoft X-ray fluxes, while 
the latter {\em increases} the nebular emission. 
%

\subsection{Wing profiles and the electron density around the WD}

Enhancement of the particle concentration around the hot star 
is supported independently by our modeling the electron-scattering 
wing profile of the \ion{O}{vi} and \ion{He}{ii} lines. 
The model parameter, the optical depth $\tau_{\rm e}$ of 
the electron-scattering layer with the thickness $r$ is 
related to its mean electron concentration $\bar{n}_{\rm e}$ 
as 
%
\begin{equation}
  \tau_{\rm e} = \sigma_{\rm T} \bar{n}_{\rm e} r .
\end{equation}
%
In spite of the model simplification (Sect.~3.2) 
we approximate the thickness $r$ of the scattering layer 
with the radius of the line emitting zone. 
Having its radius then allow us to estimate the 
$\bar{n}_{\rm e}$ quantity with the aid of Eq.~(7). 
For example, the $R_{\ion{O}{vi}}$ radius can be obtained 
from the equation of the equilibrium between the flux of 
$L_{\rm ph}(\ion{O}{v})$ photons, capable of ionizing 
\ion{O}{v} ions, and the number of recombinations by 
\ion{O}{vi} ions. By analogy with Eq.~(6), and assuming 
spherically symmetric nebular medium around the central 
ionizing source, the equilibrium condition can be expressed 
as 
%
\begin{equation}
 L_{\rm ph}(\ion{O}{v}) = \frac{4\pi}{3} 
                          R_{\ion{O}{vi}}^3\,
                          A(\ion{O}{vi})\,
                          0.83\,\bar{n}_{\rm e}^2\,
                          \alpha_{\rm B}(\ion{O}{vi}),
\end{equation}
where the factor 0.83 is the ratio of protons to electrons, 
$A(\ion{O}{vi})$ is the abundance of \ion{O}{vi} ions,
$\alpha_{\rm B}(\ion{O}{vi})$ is the recombination coefficient
of a free electron with the \ion{O}{vi} ion for the Case $B$. 
Substituting the radius of the \ion{O}{vi} zone from Eq.~(7) 
to (8), the electron concentration can be expressed as 
%
\begin{equation}
  \bar{n}_{\rm e} = 0.83\frac{4\pi}{3}\left(
        \frac{\tau_{\rm e}}{\sigma_{\rm T}}\right)^3
        A(\ion{O}{vi})\,\frac{\alpha_{\rm B}(\ion{O}{vi})}
                           {L_{\rm ph}(\ion{O}{v})} .
\end{equation}
The $L_{\rm ph}(\ion{O}{v})$ quantity as a function of 
the ionizing source temperature is showed in Fig.~B.1 
of \cite{sk+06}. 

Our model parameters, $L_{\rm h}$ and $T_{\rm h}$ from 
quiescent and transition phase (Table~3) correspond to 
$L_{\rm ph}(\ion{O}{v})$ = 4.4 and 1.7$\times 10^{44}$\,s$^{-1}$, 
respectively. 
Assuming that all oxygen atoms are ionized to \ion{O}{vi} 
within the zone, i.e. 
 $A(\ion{O}{vi}) = A(\ion{O}{}) = 4.6\times 10^{-4}$ 
\citep{aspl}, $\alpha_{\rm B}(\ion{O}{vi}) = 
 9.5\times 10^{-12}$\,cm$^{+3}$\,s$^{-1}$ \citep{gur} 
and $\tau_{\rm e}$ from Table~2, yield 
$\bar{n}_{\rm e} = 2.6\times 10^{10}$\cmt\ 
                   ($R_{\ion{O}{vi}}\dot = 50$\ro), 
and 
                  $1.7\times 10^{11}$\cmt\ 
                   ($R_{\ion{O}{vi}}\dot = 11$\ro) 
during quiescence (2004/06/15-24) and the transition 
from the burst (2003/11/14-19), respectively. 

During the optical maximum of the 2003 hot burst no 
\textsl{FUSE} observation was available. Instead, we modeled 
the \ion{He}{ii}\,$\lambda 1640$ line from the maximum of 
the 1995 hot burst (Table~1), because this line is in major 
part also created at a vicinity of the hot star. In contrast 
to the quiescent phase, an extended wing profile satisfying 
the electron-scattering broadening developed at the bottom of 
its intense emission core ($\sim 2.1\times 10^{-10}$\ecsa) 
at the optical maximum (see Fig.~2, top right). 
A relatively high optical depth, $\tau_{\rm e} = 0.32$ 
(Table~2), 
$L_{\rm ph}(\ion{He}{ii}) = 2.7\times 10^{46}$s$^{-1}$, 
$\alpha_{\rm B}(\ion{He}{ii}) =
 7.5\times 10^{-13}$\,cm$^{+3}$\,s$^{-1}$ \citep[][]{nv87} 
and 
$A(\ion{He}{ii}) = 0.1$, 
yield $\bar{n}_{\rm e} = 1.1\times 10^{12}$\cmt\
                   ($R_{\ion{He}{ii}}\dot = 6.0$\ro). 
The presence of the broad wings of 
the \ion{He}{ii}\,$\lambda 1640$ line was first pointed out by 
\cite{viotti+83} on the \textsl{IUE} high-resolution spectra 
from the 1981 outburst of AG~Dra. Simultaneously, they found 
signatures of the hot star wind suggested by the P Cygni 
profiles of the \ion{N}{v}\,$\lambda 1238,\, \lambda 1242$ 
resonance lines. 

\subsection{Flux anticorrelation and the nature of bursts}

Results of the previous two subsections are mutually 
complementary. Both the model SEDs and the profile-fitting 
analysis 
indicate a significant increase of the particle density 
around the WD during the bursts of AG~Dra. According to other 
independent analyzes 
\citep[see e.g. the \ha\ method employed by][]{sk06}, 
such the increase in the particle density is due to 
enhanced wind from the hot star into the surrounding 
particle-bounded nebula. 
This also explains the nature of this type of the optical 
bursts, because new emitters will convert the excess of 
hydrogen ionizing photons in the nebula into the nebular 
radiation that dominates the optical/near-UV. 
%
%
By return, this mechanism explains the origin of the 
X-ray--optical/UV flux anticorrelation as the result of 
variations in the wind from the hot star during different 
levels of the star's activity. 

Having the result of the reprocessing mechanism, i.e. 
the parameters of the nebular emission and/or of the wing 
profiles during bursts and quiescence, we can determine 
the mass loss rate from the accretor. Generally, the increase 
of the mass-loss rate via the wind can result from an increase 
in the mass accretion rate due to an accretion-disk instability. 
In such the case the enhanced wind provides an important 
mechanism for removal of angular momentum of rapidly accreted 
material onto the WD surface 
\citep[e.g.][]{duschl86,warner95,livio97}. 
From this point of view, the origin of the inverse flux 
correlation suggests directions for further investigation 
of the nature of the Z~And-type outbursts. 
%
%

\section{Summary and conclusions}

In this paper we investigated the origin of the supersoft 
X-ray---optical/UV anticorrelation observed during different 
levels of activity of the symbiotic binary AG~Dra. 
%

We modeled the composite continuum from the supersoft X-rays 
to near-IR during the maximum of the 2003 burst, the following 
transition to quiescence and a quiescent phase. We determined 
physical parameters of individual components of radiation 
(Fig.~2, Table~3). To support the results obtained by model 
SEDs we fitted the broad wings of the 
\ion{O}{vi}~$\lambda 1032, \lambda 1038$ and the 
\ion{He}{ii}~$\lambda 1640$ lines by the Thomson-scattering 
process. Resulting profile fits and corresponding parameters 
are shown in Fig.~2 and Table~2, respectively. 
Main results can be summarized as follows. 
\begin{enumerate}
\item
During the quiescent phase the radiation of the hot stellar 
source can be reproduced by that of a black body with a radius 
  $R_{\rm h} = 0.031 (d/1.1{\rm kpc})$\ro\ 
and radiating at the temperature 
  $T_{\rm h} = 164\,400$\,K, 
which yields the luminosity 
  $L_{\rm h} = 630 (d/1.1{\rm kpc})^2$\lo. 
The X-ray emission was attenuated by absorptions corresponding 
to the neutral hydrogen column density 
  $N_{\rm H} = 2.82\times 10^{20}$\cmd\ for the ISM abundances 
that is equivalent to its {\em interstellar} value (cf. Sect.~1). 
The emission measure of the nebular 
component of radiation was 
  $EM = 1.3\times 10^{59}(d/1.1{\rm kpc})^2$\cmt. 
Fitting parameters for the electron-scattering wings of the 
\ion{O}{vi} doublet (Table~2) and the number of 
$L_{\rm ph}(\ion{O}{v})$ photons from the model SED (Sect.~4.2) 
correspond to the mean electron concentration around the hot 
star, 
  $\bar{n}_{\rm e} \sim 2.6\times 10^{10}$\cmt. 
\item
During the transition phase these parameters changed to 
  $R_{\rm h} = 0.036 (d/1.1{\rm kpc})$\ro, 
  $T_{\rm h} = 146\,200$\,K, 
  $L_{\rm h} = 530 (d/1.1{\rm kpc})^2$\lo, 
and the absorption of the X-ray fluxes corresponded to 
  $N_{\rm H} = 3.27\times 10^{20}$\cmd. 
The larger quantity of the $N_{\rm H}$ parameter reflects 
an increase in the bound-free absorptions due to an enhanced 
wind from the hot star. 
As a result the X-ray fluxes {\em decreased} relatively to 
their values from quiescence. 
In contrast, the nebular emission {\em increased} to 
  $EM = 2.2\times 10^{59}(d/1.1{\rm kpc})^2$\cmt.
Simultaneously, flux of the broad \ion{O}{vi} wings was 
by a factor of $\sim 2$ larger than during quiescence and 
the model parameters (Tables~2 and 3, Eq.~(9)) yield 
a significantly larger density of 
  $\bar{n}_{\rm e} \sim 1.7\times 10^{11}$\cmt\ 
around the hot star, within the \ion{O}{vi} zone. 
\item
During the burst, the high limiting quantities, 
 $T_{\rm h} = 180\,000$\,K 
and 
 $L_{\rm h} = 1760\,(d/1.1{\rm kpc})^2$\lo, 
are required to produce the observed large amount of 
 $EM= 6.2\times 10^{59}(d/1.1{\rm kpc})^2$\cmt. 
The negative detection of the supersoft X-ray emission 
constraints a significant absorption effect that can 
be parametrized with $N_{\rm H} \ga 2.5\times 10^{21}$\cmd. 
The strong and extended wings of the \ion{He}{ii}\,$\lambda$1640 
line (Fig.~2) imply a large value of 
  $\bar{n}_{\rm e} \sim 1.1\times 10^{12}$\cmt\ (Eq.~(9)). 
\end{enumerate}
Based on these results we formulate the following conclusions: 

(i) 
The model SED demonstrated that the supersoft X-ray emission 
is produced by the WD photosphere. The X-ray and far-UV 
fluxes allow us to determine its temperature unambiguously. 

(ii) 
The WD's continuum spectrum is modified by the circumstellar 
and interstellar material in the line of sight. 

(iii) 
We found that the source of the opacity, causing the 
observed anticorrelation between the X-ray and optical/UV 
fluxes, can be associated with the hot star wind, which 
enhances during active phases of symbiotic binaries 
\citep[see][]{sk06}. 

(iv) 
The higher mass loss rate increases the particle density 
at the vicinity of the WD. This event increases the number 
of bound-free absorptions in the line of sight, which leads 
to a significant {\em attenuation} of the supersoft X-ray 
photons and, consequently, the free-bound transitions under 
the Case $B$ {\em increases} the nebular emission that 
dominates the optical/near-UV. 

(v)
The origin of the X-ray--optical/UV flux anticorrelation 
explains the nature of bursts in AG~Dra by reprocessing 
high-energy photons into the optical through the 
ionization/recombination events. 
Understanding the inverse relationship between optical 
and X-ray fluxes represents an important ingredient in 
the investigation of the Z~And-type outbursts.
%

\begin{acknowledgements}
This work is in part based on observations obtained with 
\textsl{XMM-Newton}, an ESA science mission with instruments 
and contributions directly funded by ESA Member States and NASA. 
The far ultraviolet data presented in this paper were obtained from
the Multimission Archive at the Space Telescope Science Institute
(MAST). They were made with the NASA-CNES-CSA Far Ultraviolet 
Spectroscopic Explorer. \textsl{FUSE} was operated for NASA by 
the Johns Hopkins University under NASA contract NAS5-32985. 
The optical spectra were obtained from the archive of the Loiano 
Station of the Bologna Atsronomical Observatory. 
This research was in part supported by a grant of the Slovak
Academy of Sciences No.~2/7010/27. The authors are grateful to 
the anonymous referee for critical comments and constructive 
suggestions. 
\end{acknowledgements}
\end{document}